\documentclass[aps,prl,preprint,superscriptaddress]{revtex4-2}
\usepackage{graphicx}

\begin{document}

\title{Charge-state resolved laser acceleration of gold ions to beyond 7 MeV/u}

\author{F. H. Lindner}
\email{florian.lindner@physik.lmu.de}
\author{E. G. Fitzpatrick}
\author{D. Haffa}
\author{L. Ponnath}
\author{A.-K.~Schmidt}
\author{M. Speicher}
\affiliation{Ludwig-Maximilians-Universität München, Am Coulombwall 1, 85748 Garching bei München, Germany}
\author{B. Zielbauer}
\affiliation{GSI Helmholtzzentrum für Schwerionenforschung GmbH, Planckstraße 1, 64291 Darmstadt, Germany}
\author{J. Schreiber}
\author{P. G. Thirolf}
\affiliation{Ludwig-Maximilians-Universität München, Am Coulombwall 1, 85748 Garching bei München, Germany}

\date{\today}

\begin{abstract}
In the past years, the interest in the laser-driven acceleration of heavy ions in the mass range \linebreak of $\text{A}\approx200$ has been increasing due to promising application ideas like the fission-fusion nuclear reaction mechanism, aiming at the production of neutron-rich isotopes relevant for the astrophysical \emph{r}-process nucleosynthesis. In this paper, we report on the laser acceleration of gold ions to beyond~7~MeV/u, exceeding for the first time an important prerequisite for this nuclear reaction scheme. Moreover, the gold ion charge states have been detected with an unprecedented resolution, which enables the separation of individual charge states up to 4 MeV/u. The recorded charge-state distributions show a remarkable dependency on the target foil thickness and differ from simulations, lacking a straight-forward explanation by the established ionization~models.

\end{abstract}


\maketitle


Promising application perspectives for laser-accelerated heavy ions in the mass range of~$\text{A}\approx 200$ led to an awakening interest in laser-based heavy ion acceleration. Since 2015, multiple experimental papers reported on progress in laser-driven acceleration of gold ions, pushing the achieved kinetic energies from~1~MeV/u~\cite{braenzel_coulomb-driven_2015} to~5~MeV/u~\cite{lindner_en-route_2019} to~finally~6.1~MeV/u~\cite{wang_production_2020}. This evolution has been accompanied by several simulations~\cite{petrov_generation_2016,petrov_heavy_2017,domanski_properties_2020}, which especially studied the expected gold ion charge-state distributions based on the established models of tunnel and electron impact ionization.

With this paper, we pursue the long-term goal of realizing the fission-fusion reaction mechanism proposed already a decade ago~\cite{habs_introducing_2011}. This aims at the production of extremely neutron-rich isotopes close to the waiting point of the rapid neutron capture (\emph{r}-)process at the magic neutron number $\text{N}=126$~\cite{arnould_r-process_2007}, which is a decisive region for the astrophysical nucleosynthesis of the heaviest elements in the Universe. The fission-fusion reaction mechanism is a two-step process, which is expected to be enabled to occur when ultra-dense bunches of laser-accelerated heavy, fissile ions (like $^{232}$Th) with kinetic energies above the fission barrier impinge on a target consisting of the same material. In a first step, both projectile and target ions undergo fission. Afterwards, fusion of fission fragments may happen, in case of fusion between two light fission fragments the desired neutron-rich \emph{r}-process isotopes are formed. This reaction scheme requires the application of laser-accelerated heavy ion bunches owing to their ultra-high, almost solid-state-like density which is expected when accelerating in the regime of radiation pressure acceleration (RPA)~\cite{esirkepov_highly_2004,macchi_laser_2005,henig_radiation-pressure_2009,kar_ion_2012}. The densities of ion bunches delivered by conventional accelerators are orders of magnitude lower and thus insufficient to allow for the realization of the fission-fusion scheme to generate these isotopes. 

Besides the still-to-be-shown acceleration of heavy ions in the RPA regime, the fission-fusion reaction process requires kinetic energies of the laser-accelerated heavy ions around~7~MeV/u (for $^{232}$Th) to exceed the fission barrier of the heavy ions. In this paper, we present for the first time experimental energy spectra from laser-accelerated heavy (gold) ions with kinetic energies that exceed this threshold of~7~MeV/u, denoting a first milestone towards the realization of this reaction mechanism. Furthermore, we show measured gold ion charge-state distributions with an unprecedented, individual-charge-state resolution. This data reveals a remarkable target thickness dependency, which lacks a straight-forward explanation by established ionization models and calls for further theoretical studies.

\begin{figure}
\includegraphics[width=0.5\textwidth]{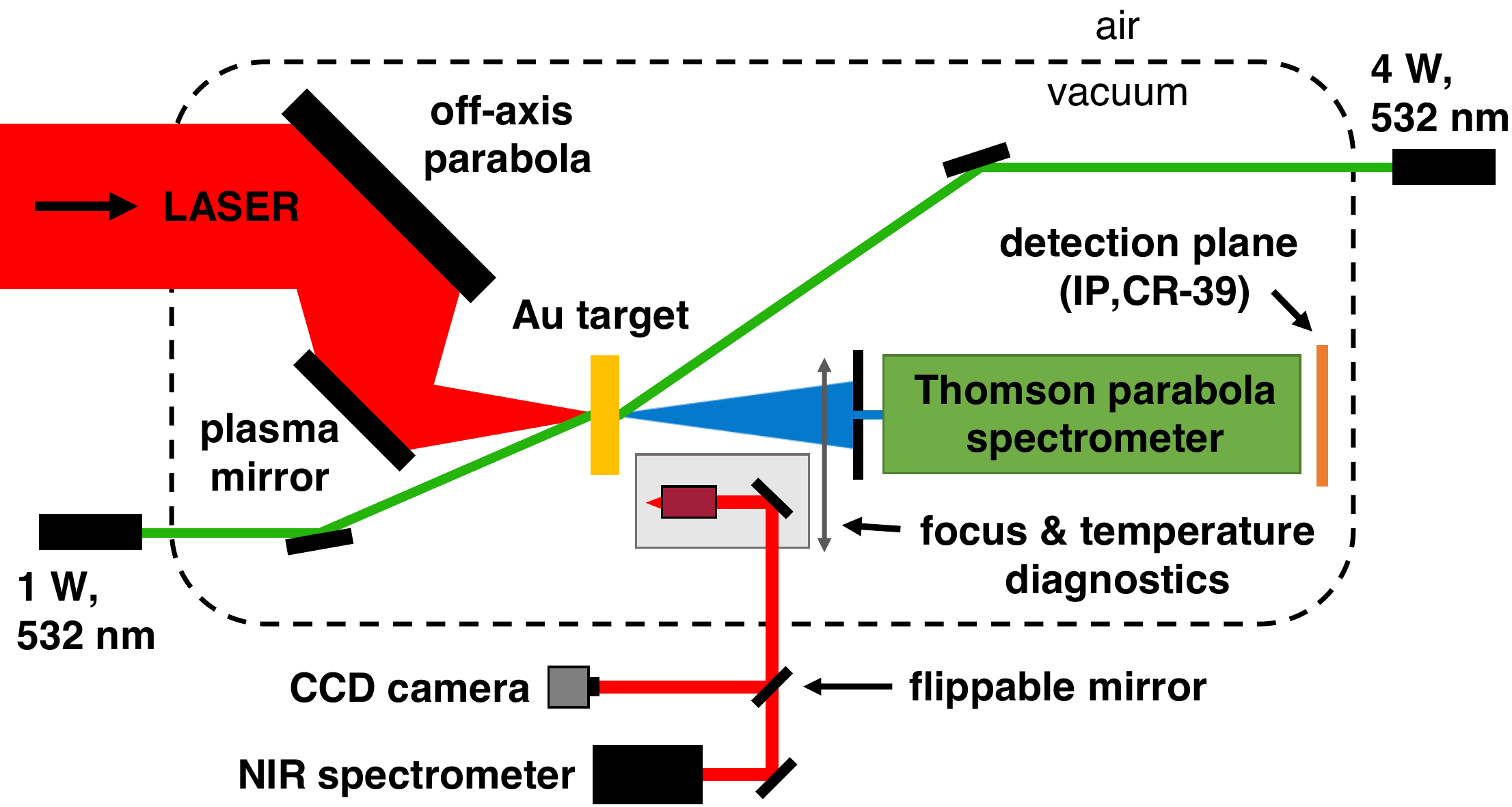}%
\caption{\label{figure1}Experimental setup at the PHELIX laser.}
\end{figure}

The experiment was conducted at the PHELIX laser of the GSI Helmholtzzentrum für Schwerionenforschung in Darmstadt, Germany~\cite{bagnoud_commissioning_2010}. A schematic overview of the experimental setup is provided in Fig.~\ref{figure1}. The PHELIX laser delivered pulses with energies of~$185 \pm 15$~J and durations of~500~fs at a central wavelength of~1054~nm. The beam was focused by an f/1.6 off-axis parabolic mirror to an area of about~$14.5 \pm 0.5$~$\mu$m$^2$, which contained around 23$\,$\% of the laser energy. UV fused silica windows were used as single inline plasma mirrors for contrast enhancement. The cycle-averaged peak intensity is estimated to be $\left(4.1\pm 0.9\right)\,\times\,10^{20}$~$\textrm{W}\,\textrm{cm}^{-2}$.

Gold targets with thicknesses of~25, 45, 100, 300 and~500~nm ($\pm10\,$\%) were provided by the LMU and GSI target laboratories. Surface contaminants were removed by radiative heating via two cw lasers at a wavelength of~532~nm (variable laser output power up to~4~W at the target front side, up to~1~W at the rear side). The thermal spectrum has been recorded outside the vacuum chamber via the optical path of the focus diagnostics by a NIR spectrometer. Although a reliable real-time determination of the foil surface temperature from the thermal spectra was not feasible during the beamtime, the targets were heated as close as possible to their melting point (1064$\,^\circ$C) by carefully increasing the laser power, while watching the thermal spectrum and the target foil surface.

A Thomson parabola spectrometer (TPS) was optimized for heavy ion characterization. The TPS features a magnetic dipole field with an average design field strength of about~850~mT along a distance of 168 mm. An electric field of up to~30 kV/cm~can be applied along a distance of up to~570~mm. 
The TPS was equipped with a stainless-steel entrance pinhole of~100~$\mu$m diameter and a thickness of 50~$\mu$m (sufficient to stop gold ions with energies up to about 12~MeV/u~\cite{ziegler_srim_2010}), collecting the ions emitted in a solid angle of $1.5\,\times\,10^{-5}$ msr in target normal direction. CR-39 sheets with dimensions of $200 \times 95 \times 1$ mm$^3$ were employed as passive ion detectors. After irradiation, they were etched in six-molar NaOH for~45~minutes at a temperature of~80$\,^\circ$C and subsequently scanned by a Zeiss Axiotron II autofocus microscope using the pattern recognition software SAMAICA~\cite{dreute_samaica_2002}. The magnetic field of the TPS was calibrated by two proton energy cutoffs at~19.92 and~22.73~MeV on imaging plates (IPs) placed behind $793\pm1$ $\mu$m and $998\pm1$~$\mu$m thick copper plates (with an average range straggling of~31 and~37~$\mu$m, respectively)~\cite{ziegler_srim_2010}. Additionally, the energy calibration was confirmed by a carbon ion energy cutoff of~18.9~MeV/u on an IP behind a~1~mm thick CR-39 sheet.

\begin{figure}
	\includegraphics[width=8.6cm]{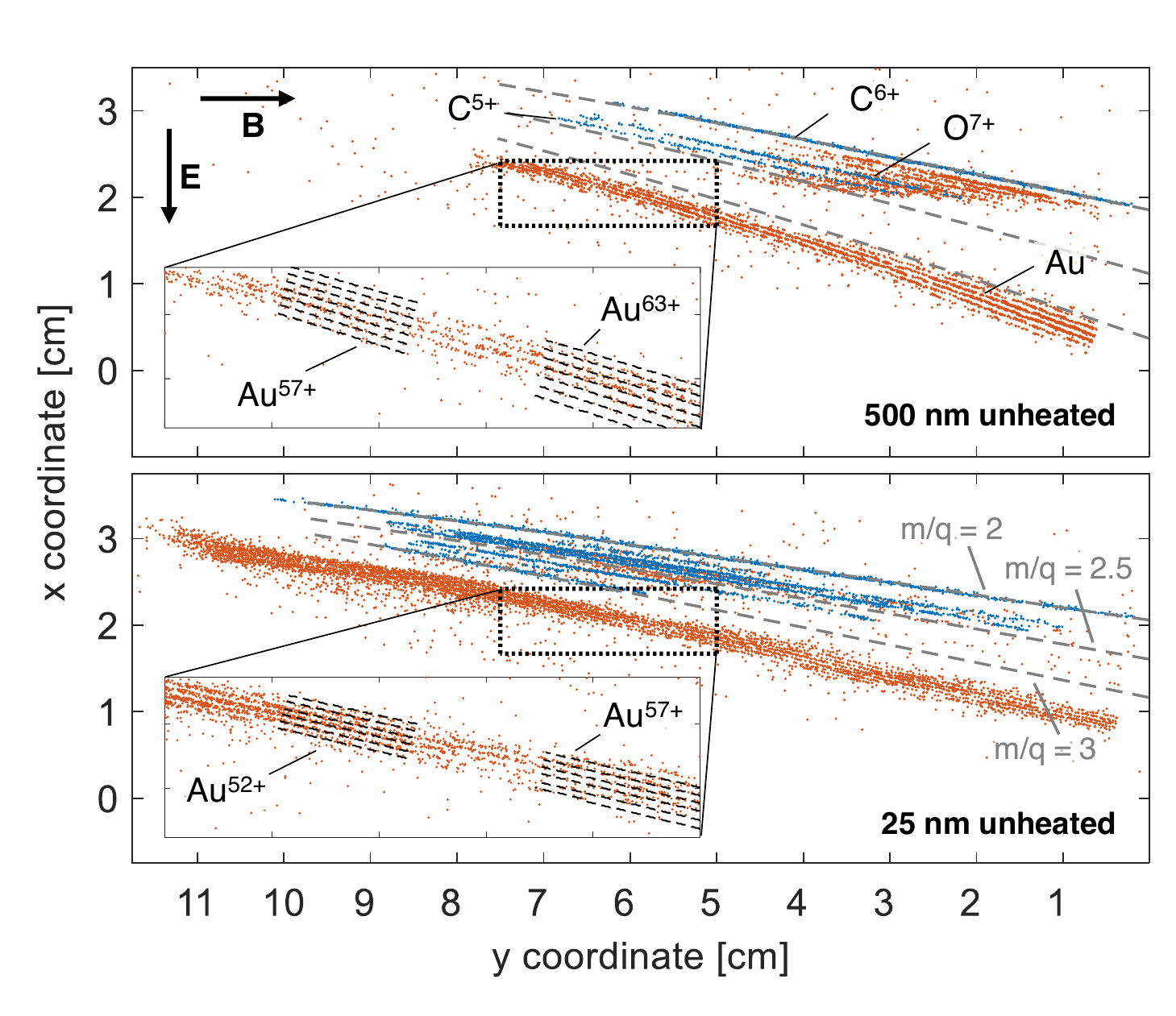}%
	\caption{\label{figure2}Raw data recorded with CR-39 for shots on unheated foils with a thickness of 500 nm (top), which delivered the best charge-state resolution and 25 nm (bottom), which resulted in the highest gold ion energies. The blue data points correspond to light ions, while the orange ones visualize larger pits caused by heavier ions, in particular gold. The grey dashed lines show the calculated lines for the mass-to-charge ratios of 2, 2.5 and~3, respectively. Details of the gold traces are shown in the insets. As indicated, the magnetic (electric) field deflected the ions to the right (downwards). }
\end{figure}

Exemplarily, the raw data from two shots on unheated gold foils with thicknesses of~25~nm (which delivered the highest gold ion energies) and~500~nm (with the best gold ion charge-state resolution) is shown in Fig.~\ref{figure2}. The ion pits could easily be identified and disentangled on the CR-39 due to their different pit sizes. The blue data points correspond to light ion impacts (carbon and oxygen) and are arranged along thin, curved traces, which can be uniquely assigned to specific mass-to-charge ratios. The orange data points originate from pits on the CR-39 with larger diameters and, thus, were associated with heavy ions, in particular gold. The gold ion impacts are also arranged along thin, curved traces, which lie much closer to each other, as expected for the smaller differences in their mass-to-charge ratio. Detailed views of the gold ion traces in the insets of Fig.~\ref{figure2} confirm the clearly distinguishable gold ion traces (below about 4~MeV/u). 
We reproducibly observed large pits in the mass-to-charge region between 2 and 2.5 (orange dots below the carbon/oxygen traces). In some shots, their faint occurance was sufficient to reveal individual, closely spaced traces. Both the large pit size and this close spacing suggest rather heavy ions as cause for this signal. Assuming consecutive charge states of a single mass number, we find the possible mass number to be between 70 and 100. It is interesting that fission fragments from a heavy-ion induced fissioning (Au + C) system after particle evaporation would qualify for this region. We note that fission fragments had been identified in PW-laser-plasma experiments~\cite{cowan_photonuclear_2000} and could in our case also originate from in-target fission caused by energetic carbon~\cite{blann_fission_1961} or oxygen~\cite{quinton_fission_1960} ions, which we also observed regularly in each shot.



\begin{figure}
	\includegraphics[width=8.6cm]{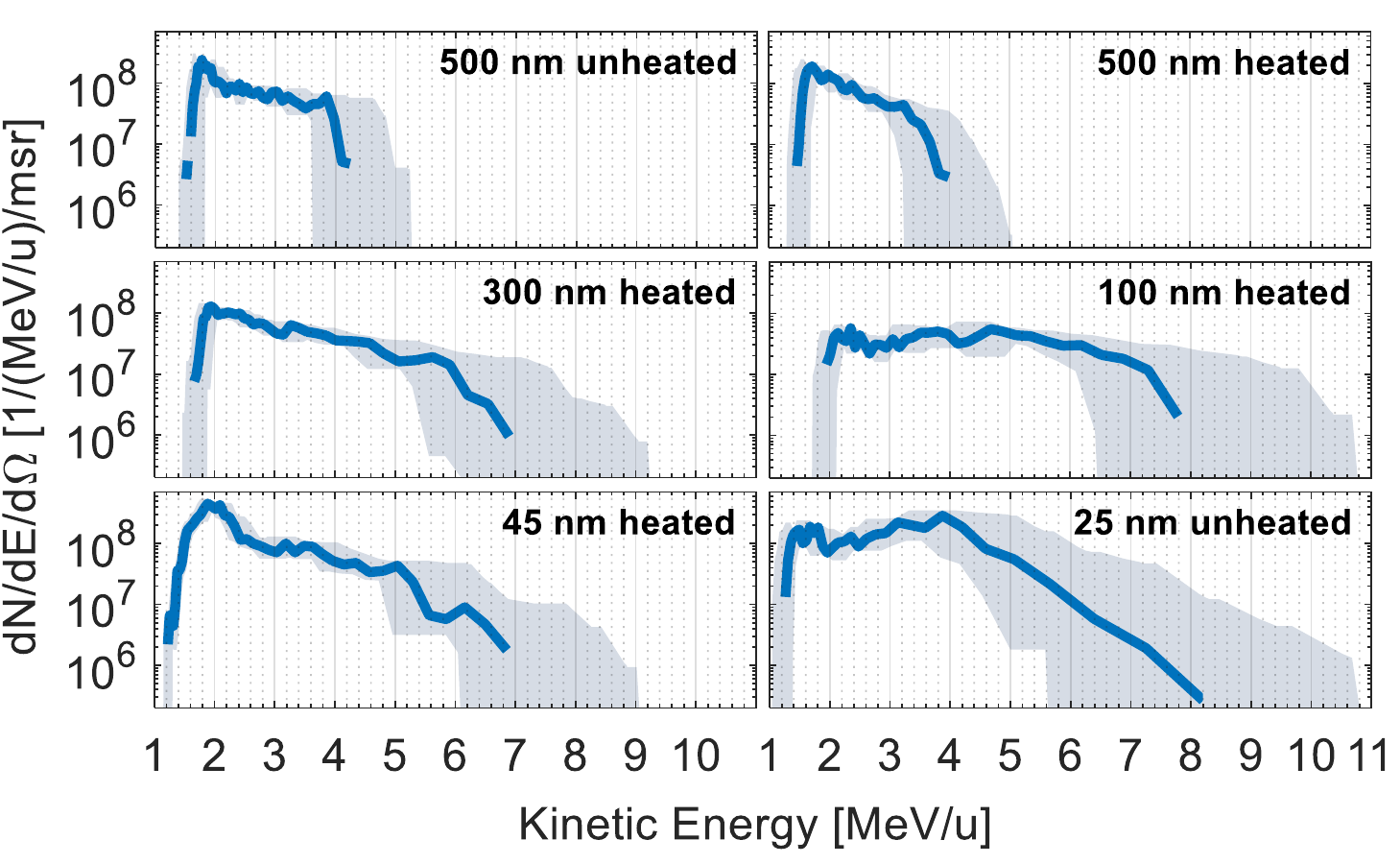}%
	\caption{\label{figure3}Gold ion energy spectra for shots on gold foils with varying thickness, integrated over all charge states. The grey error margin includes uncertainties due to the calibration of the magnetic field, the determination of the exact charge-state range and due to the intrinsic energy resolution of the TPS. Slight uncertainties of detector positioning and ion numbers have been included as well.}
\end{figure}

In total, accelerated ions from six laser shots on gold foils with varying thickness were detected on CR-39 sheets. The resulting ion spectra (integrated over all charge states) are shown in Fig.~\ref{figure3}. While from the thickest targets gold ions were accelerated up to energies around 4~MeV/u, shots on all other foils delivered reliably maximum gold ion energies above 6~MeV/u. The~best results come from the shots on 25 and 100~nm thick foils.

The target surface cleaning process with a simultaneous radiative heating of both front and rear surfaces was very effective, as the proton signal was significantly reduced for all shots on heated foils. A direct comparison is only available for the target thickness of 500 nm, where no significant effect of the heating on gold ion spectra is observed. In fact, the highest energies were achieved for a shot on an unheated, 25~nm thick gold foil. This is in contrast to former experiments~\cite{lindner_en-route_2019,hegelich_mev_2002,mckenna_characterization_2004,safronov_laser-driven_2018}. 

\begin{figure}
	\includegraphics[width=8.6cm]{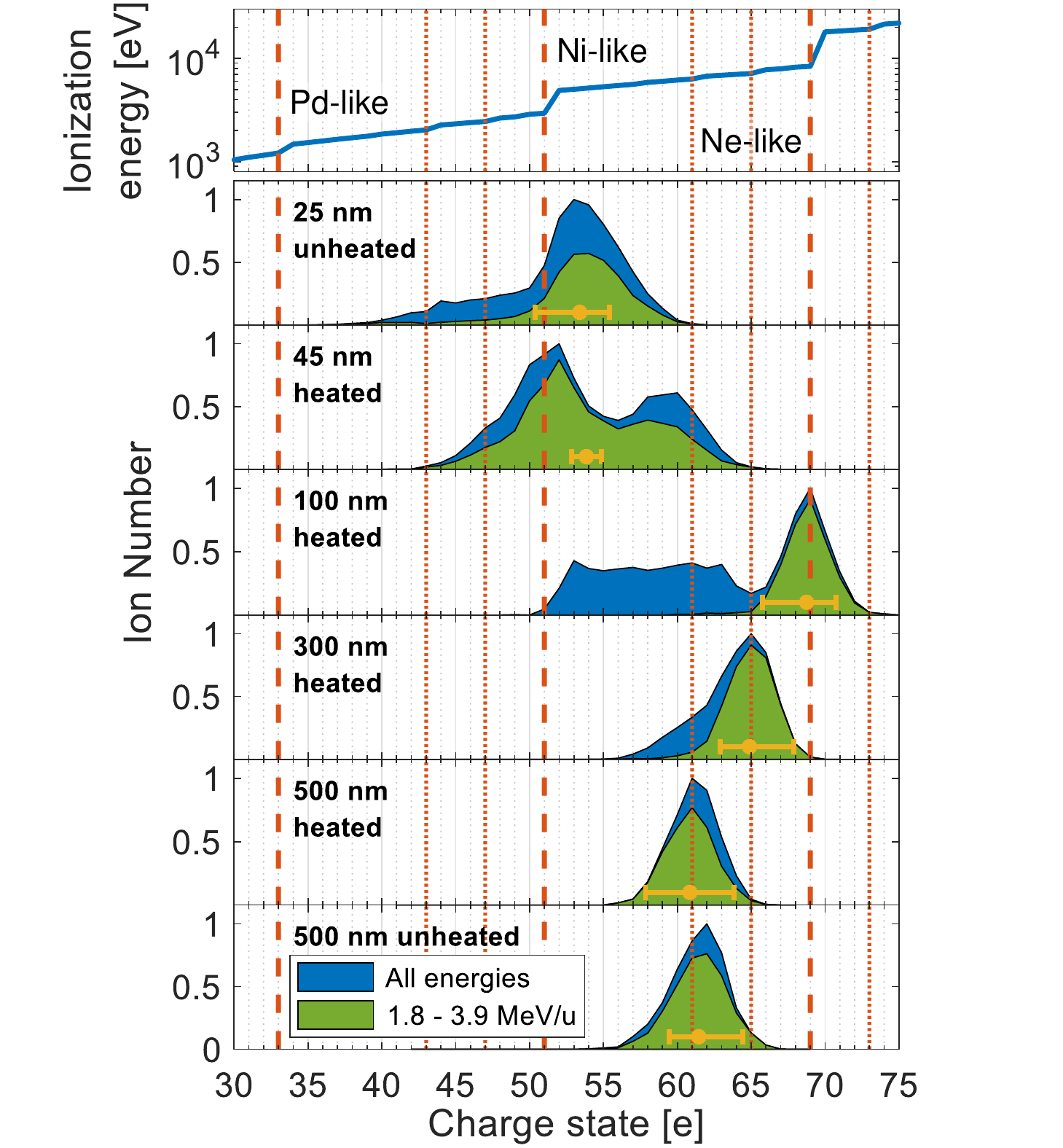}%
	\caption{\label{figure4}Comparison of the measured gold ion charge states from shots on gold foils with varying thickness to steps in the gold ionization energy indicated by the vertical orange lines (dashed lines correspond to larger, dotted lines to smaller steps). The blue distributions show the gold ion numbers for each charge state integrated over all energies. The green distributions display the number of gold ions integrated between 1.8 and 3.9 MeV/u, for which the individual charge states were resolvable for most of the shots. The yellow point depicts the mean charge state of the green distribution with an error bar showing the uncertainty of the	total charge state range. The distributions are normalized to the respective maximum of the blue curves.}
\end{figure}

Besides these encouraging kinetic energies, we detected the laser-accelerated gold ions with an unprecedented charge-state resolution, which enabled us for the first time to resolve individual charge states up to energies of about~4~MeV/u. The measured gold ion charge-state distributions, normalized to their maximum value, are plotted in Fig.~\ref{figure4}, integrated over all energies (blue) and within an energy range between 1.8 and 3.9 MeV/u (green), where individual charge states could be resolved for most of the shots. It is obvious that the width of the charge-state distributions increases towards thinner targets. The TPS allowed measuring charge states higher than $(36^{+2}_{-3})^+$ for a foil thickness of~25~nm, while the highest measured charge state is $(74^{+2}_{-3})^+$ from a~100~nm thick gold foil, which reaches the recently published record ionization value of $72^+$, representing the highest charge state of gold that has been observed so far~\cite{hollinger_extreme_2020}. A significant difference between the charge-state distributions from heated and unheated foils was not observed. 

It is striking that each of the charge-state distributions -- even the broad ones for thinner target foils -- exhibits a clear maximum. Prominent peaks are also visible at individual charge states in the respective distributions shown in the simulation papers~\cite{petrov_generation_2016,petrov_heavy_2017,domanski_properties_2020}, with a predominant occupation of the charge state 51$^+$. These peaks can be attributed to large steps in the ionization potential, which is shown for the sequential ionization of gold atoms in the topmost panel of Fig.~\ref{figure4} \cite{kramida_nist_2019}. The thick, dashed orange lines mark positions in the charge-state spectra from major steps in the ionization potential that occur for closed atomic shells: 3$3^+$ is palladium-like with a closed 4d shell, 51$^+$ is nickel-like with a closed 3d shell and 69$^+$ is like the noble gas neon with a closed 2p shell. The thinner, dotted orange lines visualize steps in the ionization potential with step sizes that are at least a factor of 2 higher than neighbouring steps, which result mostly from closed subshells as well. 

However, despite the single-charge-state resolution, the peaks in the measured distributions are not as sharp as in the simulations, where distinct peaks stand out compared to the neighbouring charge states. Instead, the measured maxima are broadened and distributed over a width of at least three charge states. Nevertheless, these peaks coincide -- within their uncertainties for the charge-state range -- very well with the positions of the major and minor steps in the ionization potential and show a remarkable thickness dependency: the charge states from thinner foils ($<$~50~nm) peak around 51$^+$ with a rather broad underlying charge-state distribution, while gold ions from thicker foils are quite closely spread around maxima at much higher charge states. The distributions for the two shots on~500~nm thick gold foils are in excellent agreement with each other and the highest populated charge state lies around~61$^+$ for both cases. For decreasing foil thicknesses, the widths of the charge-state distributions increase and their peak positions move towards higher values, all coinciding with major and minor steps in the ionization energies (around~65$^+$ for~300~nm and~69$^+$ for~100~nm). At a gold foil thickness of~100~nm, a uniform distribution at lower charge states between~51$^+$ and~65$^+$ arises in addition to the peak at~69$^+$. This appears to be the onset of a transition towards the charge-state spectra from thinner foils with thicknesses of~45~and~25~nm, which are relatively broad compared to the thicker foils and located at much lower charge states.

A straight-forward explanation for the measured gold ion charge-state distributions and especially the observed foil thickness dependency could not be found in the framework of established ionization mechanisms (tunnel and electron impact ionization). Regarding the tunnel ionization, the established formulae applying the ADK-model~\cite{ammosov_tunnel_1986,chang_closed-form_1993} predict for the here measured peak intensity of $\left(4.1\pm 0.9\right)\,\times\,10^{20}$ $\textrm{W}\,\textrm{cm}^{-2}$ the ionization of gold up to~51$^+$, which corresponds to the large step in the ionization potential for a nickel-like configuration. However, the cycle-averaged peak intensity is naturally a factor of two lower than the instantaneous peak value, which acts on the ions in the plasma. Additionally considering the long laser pulse duration, it is possible that the target foil surface expands and becomes relativistically transparent before the laser pulse has ended. In this case, the laser penetrates the foil until it faces deeper target layers with densities above the critical value, at which the laser is reflected back. This would further enhance the intensity by a factor of 4, yielding a peak value of about $3\,\times\,10^{21}$ $\textrm{W}\,\textrm{cm}^{-2}$, which would already be sufficient to ionize gold up to~61$^+$ to~65$^+$. The occurrence of relativistic self-focusing is conceivable as well~\cite{bin_ion_2015}, which would considerably increase the laser intensity and the generation of charge states of 69$^+$ (requiring an intensity of $5\,\times\,10^{21}$ $\textrm{W}\,\textrm{cm}^{-2}$) and even above becomes viable.

For the assessment of the relevance of collisional ionization, the Lotz formula \cite{lotz_empirical_1967} has been used with the ionization potentials from reference \cite{kramida_nist_2019} as input instead of the binding energies. Assuming a gold foil thickness of 500 nm, we derive a total electron number of $N = Z\cdot4.4\times10^{11}$ and end up with a probability of about $4\,\%$ for the ionization from $51^+$ to $52^+$ and of about $0.6\,\%$ for the ionization from $69^+$ to $70^+$. The influence of a possible return current from regions surrounding the focal spot as discussed in \cite{nishiuchi_dynamics_2020}, which could potentiate the number of contributing electrons, in particular at low kinetic energies, and thus the ionization probabilities, has not yet been considered.

\begin{figure}
	\includegraphics[width=8.6cm]{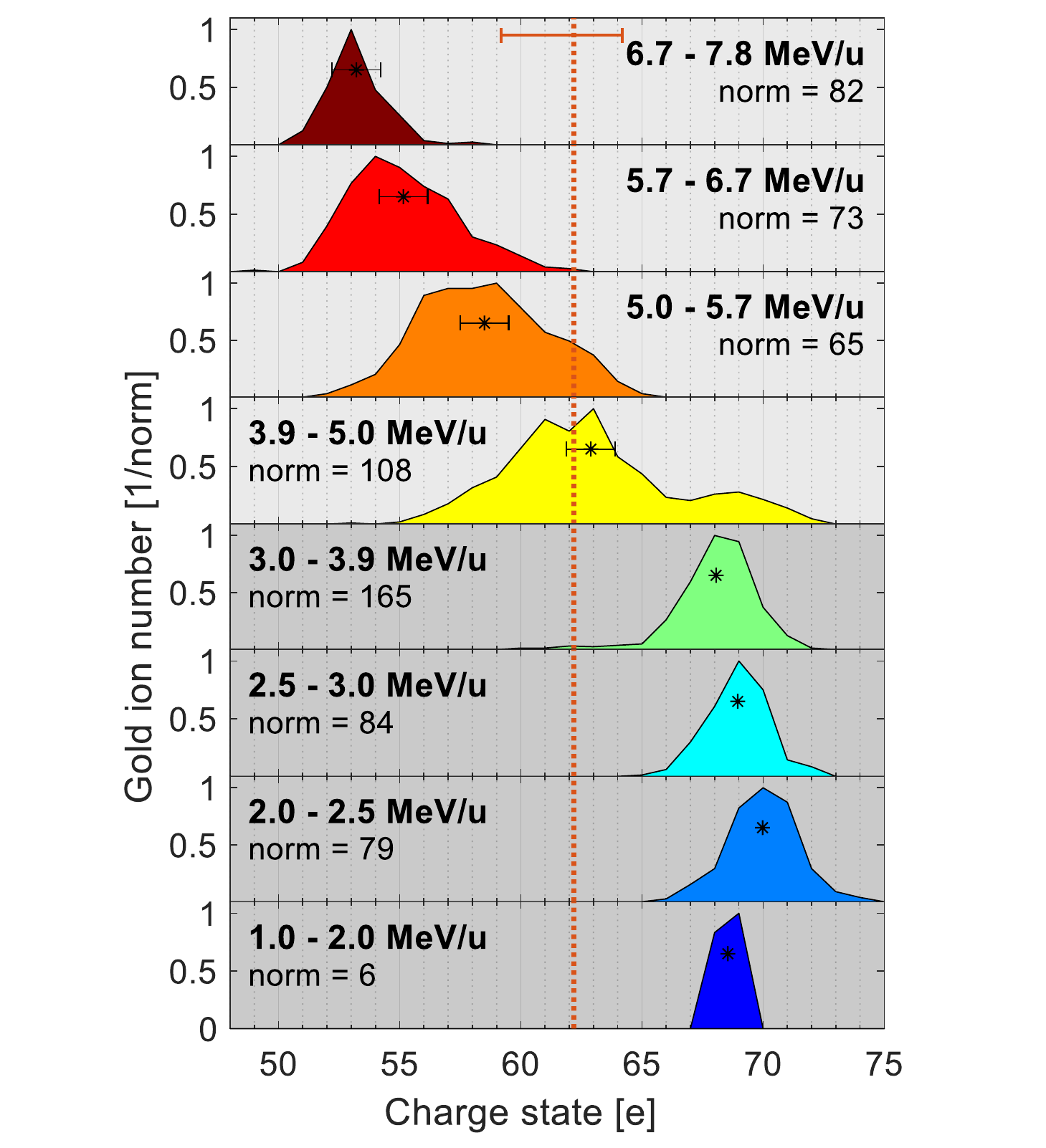}%
	\caption{\label{figure5}Energy-dependent gold ion charge-state distributions for a shot on a heated, 100~nm thick gold foil. The energies are increasing from the bottom to the top, which is also visualized by the face colours of the distributions (blue means low, red means high energy). The ion numbers of each distribution were normalized to their respective maximum, which is stated for each panel individually ('norm~=~X'). The respective mean value has been marked by an asterisk (*) for each charge-state distribution. The error bars in the four topmost panels indicate the charge-state resolution within the shown distribution. The total charge state mean value, averaged over all kinetic energies, is indicated by the dotted, orange line. In the topmost plot, the uncertainty of the charge-state range itself is indicated by the orange error bar.}
\end{figure}

Fig.~\ref{figure5} shows the energy-resolved gold ion charge-state distribution, exemplarily for a shot on a heated, 100 nm thick gold foil. The kinetic gold ion energies are increasing from the bottom towards the top. The respective energy intervals are stated in each panel. It is striking that the gold ion charge states decrease with increasing kinetic energy, especially for energies exceeding~4~MeV/u. At a first glance, this appears counterintuitive, as ions with lower mass-to-charge ratio are generally exposed to higher acceleration forces. The authors of reference \cite{domanski_properties_2020} have observed a similar behaviour in their simulations. They explained their finding by the sequential nature of the ionization dynamics in combination with the relatively long laser pulse: after an initially similar degree of ionization, faster particles leave the high-ionizing-field area relatively early and thus keep their lower charge state during their acceleration phase, while slower particles remain in the high-field region, being further ionized to higher charge states. Despite their more efficient acceleration, these particles cannot catch up with the lowly charged particles due to the limited remaining acceleration time. 

In conclusion, we have detected laser-accelerated heavy (gold) ions with energies exceeding 7 MeV/u, which represents the demonstration of an important milestone towards the realization of the novel fission-fusion reaction mechanism. Assuming zero emittance and based on the absolute particle numbers from our analysis, we estimate the ion bunch density at this energy to be around $10^{13}\,\frac{1}{\text{cm}^3}$ ($10^{16}\,\frac{1}{\text{cm}^3}$) at a distance of 1 mm (100 $\mu$m) from the target. These promising high densities motivate further studies on potential collective effects on the ion stopping or energy loss behaviour in solid or gaseous media at even higher laser intensities. The high charge state resolution in our experiments provide novel, experimental data which challenges theoretical models that rely on the two most commonly used ionization models. Although these processes have the potential to ionize gold atoms up to the measured charge states, neither the remarkable target foil thickness dependency nor the narrow width of the charge-state distributions for thicker gold foils can be explained in a straight-forward and intuitive way. Therefore, this data constitutes a valuable input for further theoretical investigations targeting to analyze the exact contributions of different ionization mechanisms in laser-generated plasmas.

\begin{acknowledgments}
	The results presented here are based on the experiment P174, which was performed at the PHELIX facility at the GSI Helmholtzzentrum fuer Schwerionenforschung, Darmstadt (Germany) in the frame of FAIR Phase-0. This work was funded by the BMBF under contract 05P18WMEN9, the DFG Cluster of Excellence MAP (Munich-Centre for Advanced Photonics) and the Centre for Advanced Laser Applications (CALA) in Garching. AKS acknowledges support from DFG GRK2274, DH from BMBF under contract 05P18WMFA1. The authors acknowledge the PHELIX team for their support during the experiment and the LMU and GSI target laboratories for preparing the target foils.
	\end{acknowledgments}
\bibliography{Bibliothek.bib}

\end{document}